# Extreme fast charging of batteries using thermal switching and self-heating

Yuqiang Zeng[1,†], Buyi Zhang[1,2,†], Yanbao Fu[1], Fengyu Shen[1], Qiye Zheng[1,2], Divya Chalise[1,2], Ruijiao Miao[1,2], Sumanjeet Kaur[1], Sean D. Lubner[1], Michael C. Tucker[1], Vince Battaglia[1], Chris Dames[1,2], and Ravi S. Prasher[1,2,*]

[1]Energy Storage and Distributed Resources Division, Lawrence Berkeley National Laboratory, Berkeley, CA, 94720, USA

[2]Department of Mechanical Engineering, University of California, Berkeley, CA, 94720, USA

[†]: These authors contributed equally.

[*]: Author to whom correspondence should be addressed.

Email: rsprasher@lbl.gov

**The long charge time of electric vehicles compared with the refueling time of gasoline vehicles, has been a major barrier to the mass adoption of EVs[1–4]. Currently, the charge time to 80% state of charge in electric vehicles such as Tesla with fast charging capabilities is >30 minutes[5]. For a comparable recharging experience as gasoline vehicles, governments and automobile companies have set <15 min with 500 cycles as the goal for extreme fast charging (XFC) of electric vehicles[6,7]. One of the biggest challenges to enable XFC for lithium-ion batteries (LIBs) is to avoid lithium plating[8–10]. Although significant research is taking place to enable XFC[11–14], no promising technology/strategy has still emerged for mainstream commercial LIBs. Here, we propose a thermally modulated charging protocol (TMCP) by active thermal switching for XFC, *i.e.*, retaining the battery heat during XFC with the switch OFF for boosting the kinetics to avoid lithium plating while dissipating the heat after XFC with the switch ON for reducing side reactions. Our proposed TMCP strategy**

**enables XFC of commercial high-energy-density LIBs with charge time <15 min and >500 cycles while simultaneously beating other targets set by US Department of energy[6] (discharge energy density > 180 Wh/kg and capacity loss <4.5%). Further, we develop a thermal switch based on shape memory alloy and demonstrate the feasibility of integrating our TMCP in commercial battery thermal management system**.

It is widely acknowledged that long XFC cycle life cannot be achieved in existing commercial high-energy-density lithium-ion batteries (LIBs) with graphite (C) anodes and transition metal oxide cathodes such as lithium cobalt oxide (LCO)[2]. Reducing the charge time to 15 min needs a charge rate of 6C, which can trigger lithium plating on graphite anodes and cause dramatic capacity fade in LIBs. Eliminating or mitigating lithium plating[8–10], which requires faster ion transport and kinetics in LIBs, is one of the biggest R&D challenges for enabling XFC. Broadly, R&D efforts to develop XFC LIBs can be categorized as the development of new electrolytes[11,15], electrode materials[12,16,17], charge protocols[13,18], and heating strategies (*i.e.*, improving the kinetics by increasing the temperature before XFC)[14,19–21]. Among these approaches, only heating strategies have shown promising results for existing high-energy-density LIBs and thus have the potential to enable XFC of electric vehicles in the near term.

With the heating strategy, the battery temperature ($T_B$) is modulated as $T_B \approx Q/(hA) + T_C$, where $Q$ is the heat generated in the battery, $T_C$ is the coolant temperature, $A$ is the surface area of the battery, and $h$ denotes the tunable thermal conductance per unit area between the battery and the coolant as shown in Fig. 1a. Two heating strategies have been proposed/enacted for fast charging: 1) System-level $T_B$ control using battery thermal management systems (BTMSs)[20,21] by adjusting $h$ and $T_C$ using coolant modulation, *e.g.*, increasing or reducing $h$ by starting or stopping the coolant flow and/or changing $T_C$ by heating or cooling the coolant. $T_B$ is raised by reducing $h$ and/or increasing $T_C$ during fast charging and reduced during rest and discharge by increasing $h$ and/or reducing $T_C$. In fact, coolant-controlled charge protocols are being adopted by EV companies[20], however, the maximum charge rate is only 2C as opposed to the possibility of 4C-6C suggested by electrochemical-thermal (ECT) simulations[21]. The difference in rate capability between the simulation and real world is from the low gravimetric (0.55-0.65) and volumetric (<0.4) cell-to-pack (CTP) ratio in practical battery packs[22]. Based on our validated ECT model of battery packs with such low CTP ratios (Supplementary Fig. 1 and Note 1), a large portion of battery heat (~40%) is dissipated to the pack due to its high thermal mass even if coolant flow is completely stopped, thereby limiting the temperature rise during charging. This also results in negative anode potential which indicates high likelihood of lithium plating. 2) cell-level $T_B$ control with increased $Q$ using embedded nickel foil heaters and reduced $h$ by thermally insulating the cell[14]. This method enables higher C rates as the increase of $T_B$ is observably higher than the system-level strategy. However, there are two major challenges with this method. Since the battery is always thermally insulated, $T_B$ is high even during discharge and rest and the high average $T_B$ during operation can dramatically lower the overall performance and lifetime[23–25] (Supplementary Fig. 2). This effect becomes noticeable with increasing ambient temperature as the side reaction rates increase rapidly with the temperature. Further, adding extra metal foils in the cell is incompatible with existing battery manufacturing process which has been perfected by the industry over decades. The intrusive nature of the embedded heater may also raise safety concerns.

In this work, we propose a thermally modulated charging protocol (TMCP) which combines the good traits of both the cell and system-level strategies by active thermal switching (ATS), *i.e.*, low $h$ ($h_{off}$) at cell level during XFC and high $h$ ($h_{on}$) during discharge/rest (Fig. 1a). A proof-of-concept study of commercial high-energy-density LCO/C LIBs under various thermal protocols demonstrates that our TMCP consistently outperforms existing thermal protocols. With the TMCP, the XFC performance of commercial LIBs exceeds the key DOE targets. Based on electrochemical analysis and postmortem characterizations using optical microscopy, scanning electron microscopy (SEM), and X-ray tomography, the improved XFC performance is attributed to the mitigated lithium plating during XFC and reduced side reactions during discharging by ATS. For practical implementation of our approach with existing BTMSs, we have developed an ATS device with small mass and volume (1.4% and 3.0% compared to that of battery, respectively) using cost-effective shape memory alloy, which has the potential to enable XFC in commercial battery packs.

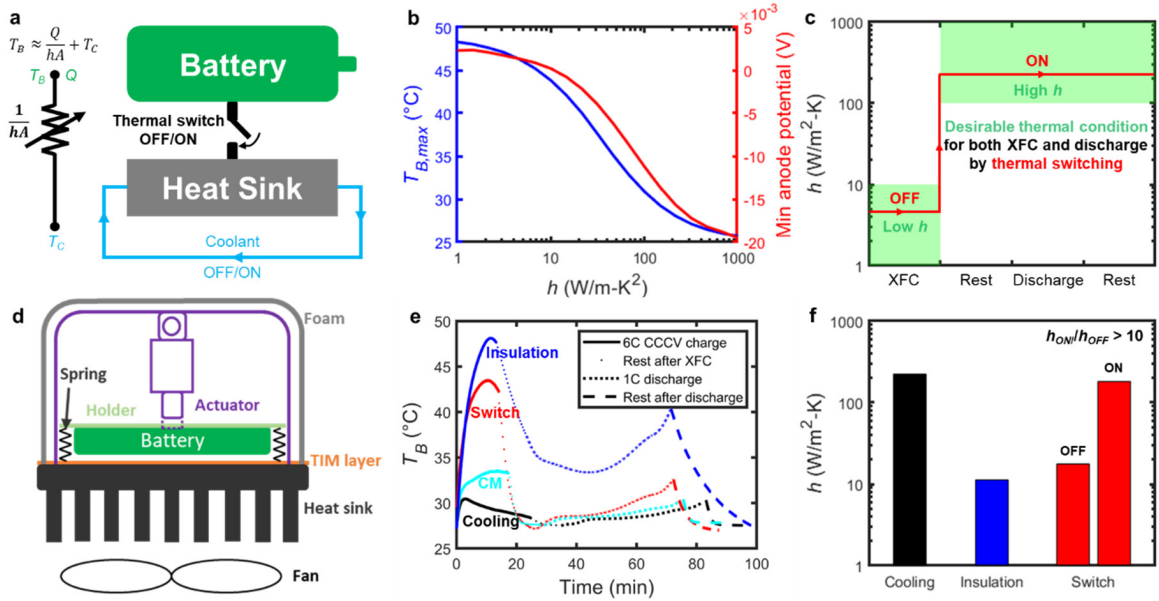

**Fig. 1 | Active thermal switching for XFC. a,** Schematic of thermal switching concept for retaining the heat during XFC with the switch OFF and dissipating the heat during resting and discharging with the switch ON. **b,** The maximum battery temperature and the minimum anode potential during XFC as a function of thermal conductance per unit area *(h)* for the cell. The minimum anode potential is positive during XFC when the cell is thermally insulated with $h \leq 10$ W/m²-K. **c,** Thermally modulated charge protocol (TMCP) with thermal switching ratio >10 for XFC. Note in the OFF state of the thermal switch, the coolant flow is also off whereas during ON state coolant flow is also on. **d,** Linear actuator to mimic ATS for conducting controlled experiments. The gap between the battery and heat sink can be tuned as the actuator contracts or elongates and thus the thermal contact changes. **e,** Representative battery temperature evolution in a XFC cycle with different thermal protocols. **f,** Effective *h* between the battery and the coolant for various cases obtained by matching thermal model with the experimental data. *h* with the switch OFF and ON is comparable to *h* in the case of insulation and cooling, respectively.

To design the TMCP, we evaluate the impact of *h* on XFC of representative LCO/C cells using our ECT model (Supplementary Fig. 1 and Note 1). The simulation indicates the range of $h_{off}$ needed to reach optimal $T_B$ and avoid negative anode potential for lithium plating mitigation during XFC (Fig. 1b). Figure 1c reveals that a switching ratio ($h_{on}/h_{off}$) of ~10 is needed as high $h_{on}$ is desired for the other states. As a proof of concept, we perform 6C1C cycling tests (6C charge and 1C discharge; see Methods for details) of commercial high-energy-density LIBs with our TMCP. Representative 5Ah LCO/C LIBs with C/3 energy density 240.8 Wh/kg, maximal charge rate 1C, and maximum discharge rate 2C are used for this XFC study (see details in Methods). A linear actuator is used to simulate ATS experimentally to validate the efficacy of TMCP (Fig. 1d and Supplementary Fig. 3). The springs hold the battery during XFC (switch OFF and coolant flow off), while the actuator elongates after XFC and pushes the battery in contact with the heat sink (switch ON and coolant flow on). Apart from ATS, controlled experiments are conducted with other thermal protocols for XFC: 1) cooling: the coolant flow is always on (*i.e.*, the conventional thermal scheme[26]); 2) coolant modulation (CM OFF/ON): the coolant flow is off during XFC and on during resting and discharging (*i.e.*, the system-level strategy[18]); and 3) insulation: the cell is always thermally insulated as proposed by Yang *et al.*[14]. Fig. 1e displays the representative battery temperature evolution in a XFC cycle under different thermal protocols. With the switch OFF, the rise of $T_B$ during XFC is comparable to the insulation case. After XFC, turning on the switch allows for efficient cooling and optimal control of $T_B$. In contrast, the insulation case leads to high temperature during rest and discharge which is highly undesirable[25] (Supplementary Fig. 2). By fitting the temperature profile with thermal models (Supplementary Fig. 4 and Note 2), the effective $h$ is extracted for various cases (Fig. 1f). It shows that the switching ratio of 10.4 is >10 as needed (Fig. 1c) and the effective $h$ with the switch OFF and ON is comparable to $h$ in the case of insulation and cooling, respectively.

For this type of cell, the charge time ($t_c$) to 80% state of charge (SOC) depends highly on the thermal protocol and temperature rise (Fig. 2a and Supplementary Fig. 5). The charge time with the coolant flow on (cooling) and off (CM OFF) is ~25 min and ~18 min, respectively. It decreases to <15 min in the case of insulation and switch due to the boost of battery kinetics by high temperature. As discussed earlier, the high temperature is beneficial to avoid negative anode potential and mitigate Li plating during XFC, which can be verified using the voltage feature related to the intercalation of plated Li during rest[27–29]. We calculate the derivative of voltage ($V$) with respect to time ($t$), d$V$/d$t$, from the resting voltage after XFC (Fig. 2b). An inflection point feature is observed in the plot of d$V$/d$t$ for the case of cooling and CM, while no such feature appears in the case of insulation and switch. The disappearance of the inflection feature indicates the mitigation of lithium plating and explains the high coulombic efficiency of these two cases (Supplementary Fig. 6). This further leads to the extended 6C1C cycle life associated with 20% capacity fade (Fig. 2c). The lower cycle life in the insulation case (665), compared to the switch (975), is attributed to the increased side reaction rates at higher discharge temperature. This effect is more pronounced at higher ambient temperature (~40 °C) resulting in cycle life of 334 for the insulation as compared to 560 for the switch which still exceeds the US DOE target (Figs. 2d and 2e). Figure 2f demonstrates the XFC cycle life which is defined as the number of cycles that batteries can be charged to 80% SOC in 15 min. Due to the limit of $t_c$ (Supplementary Fig. 5), the XFC cycle life can be lower than the overall cycle life at 20% capacity loss (Fig. 2b), *e.g.*, the XFC cycle life for the case of cooling and CM is 0. Our approach enables >500 XFC cycles at different ambient temperatures and the superiority of "switch" over "insulation" increases with ambient temperature. Further, we perform 1C1C cycling tests (1C charge and 1C discharge) using normal cooling condition and compare the results with those of 6C1C cycling using the ATS (Supplementary Fig. 7). The comparable operation time of 6C1C and 1C1C cycling reveals that the degradation related to XFC is largely reduced with our TMCP.

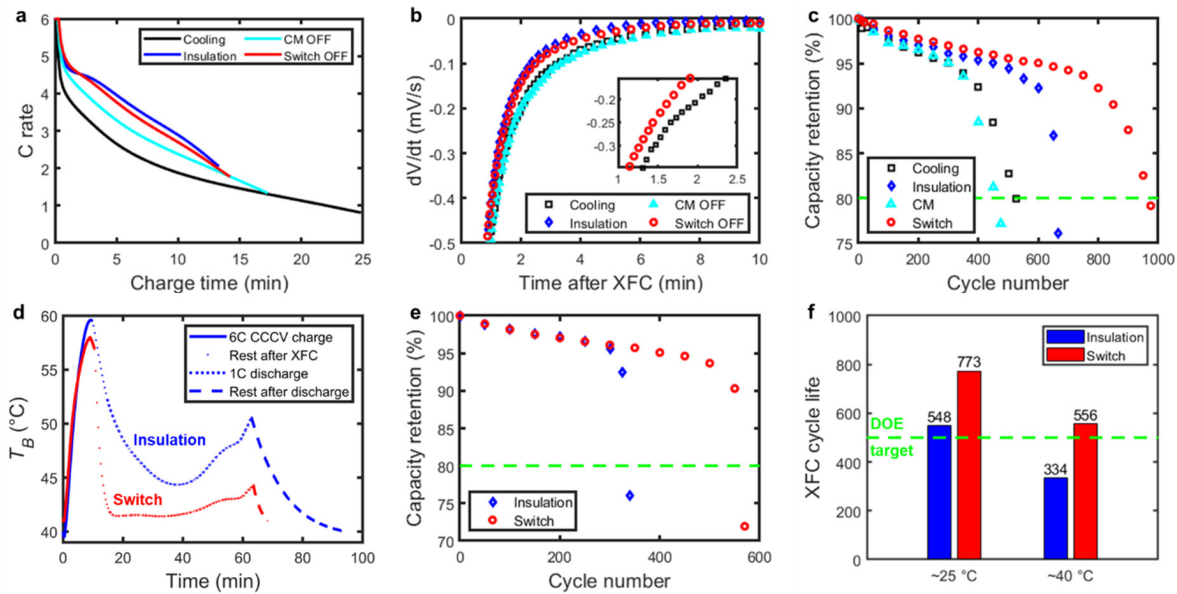

**Fig. 2 | XFC cycling results. a,** Evolution of C rate during 6C CCCV charging to 80% SOC. **b,** The derivative of resting voltage after XFC with respect to time. The inflection point feature observed in the case of cooling and CM confirms the severe lithium plating during XFC. **c,** Capacity retention of the cells cycled at 25±5 °C with different thermal protocols. **d,** Representative temperature evolution for the cells tested at ~40 °C ambient temperature in the case of insulation and switch, and the corresponding capacity retention in **e**. The long-time exposure to >45 °C leads to an early degradation in the case of insulation. **f,** XFC cycle life of the cells tested at ~25 °C and ~40 °C. Our approach enables >500 XFC cycles and exceeds the US DOE target at different ambient temperatures.

Capacity loss behavior shown in Fig. 2c is due to complex combination of Li plating and side reactions such as solid electrolyte interphase (SEI) growth[30–32]. To better understand the capacity fade in the case of ATS and cooling (*i.e.*, our TMCP and the conventional thermal scheme for XFC), we investigated the degradation mechanism with electrochemical analysis (Fig. 2b, Supplementary Fig. 8) and postmortem characterizations using optical microscopy, SEM, and X-ray tomography (Fig. 3). In the case of cooling, a large portion of the aged anode is covered by plated Li, and the particle feature becomes hardly visible due to the coverage (Fig. 3b and 3e). By comparison, the particle feature remains visible in most part of the aged anode while some particles are covered by a layer of reaction products in the case of ATS (Fig. 3c and 3f). These observations confirm the presence (cooling) or mitigation (ATS) of lithium plating as revealed by the *dV/dt* analysis (Fig. 2b). This explains the different rate of capacity fade in the initial linear aging regime, which largely determines the cycle life. For the rapidly increased capacity loss after the initial stage, such transition has been reported in prior works[30–32] and is typically attributed to the rapid increase of lithium plating associated with battery aging, *e.g.*, electrolyte consumption and reduced anode porosity. The reduction of porosity in the aged anodes is quantified with tomography (Fig. 3g-i). Compared to the cooling case, the higher porosity loss in the aged anode from the ATS case is in part a result of the more SEI growth related to the longer cycle life and higher operation temperature. This agrees with the larger SEI resistance observed in the aged cell from "ATS"(Supplementary Fig. 8 and Table 1). Lastly, we confirm that the loss of active materials or change of cathode has a limited impact on the observed capacity degradation (Supplementary Table 2 and Supplementary Fig. 9 & 10).

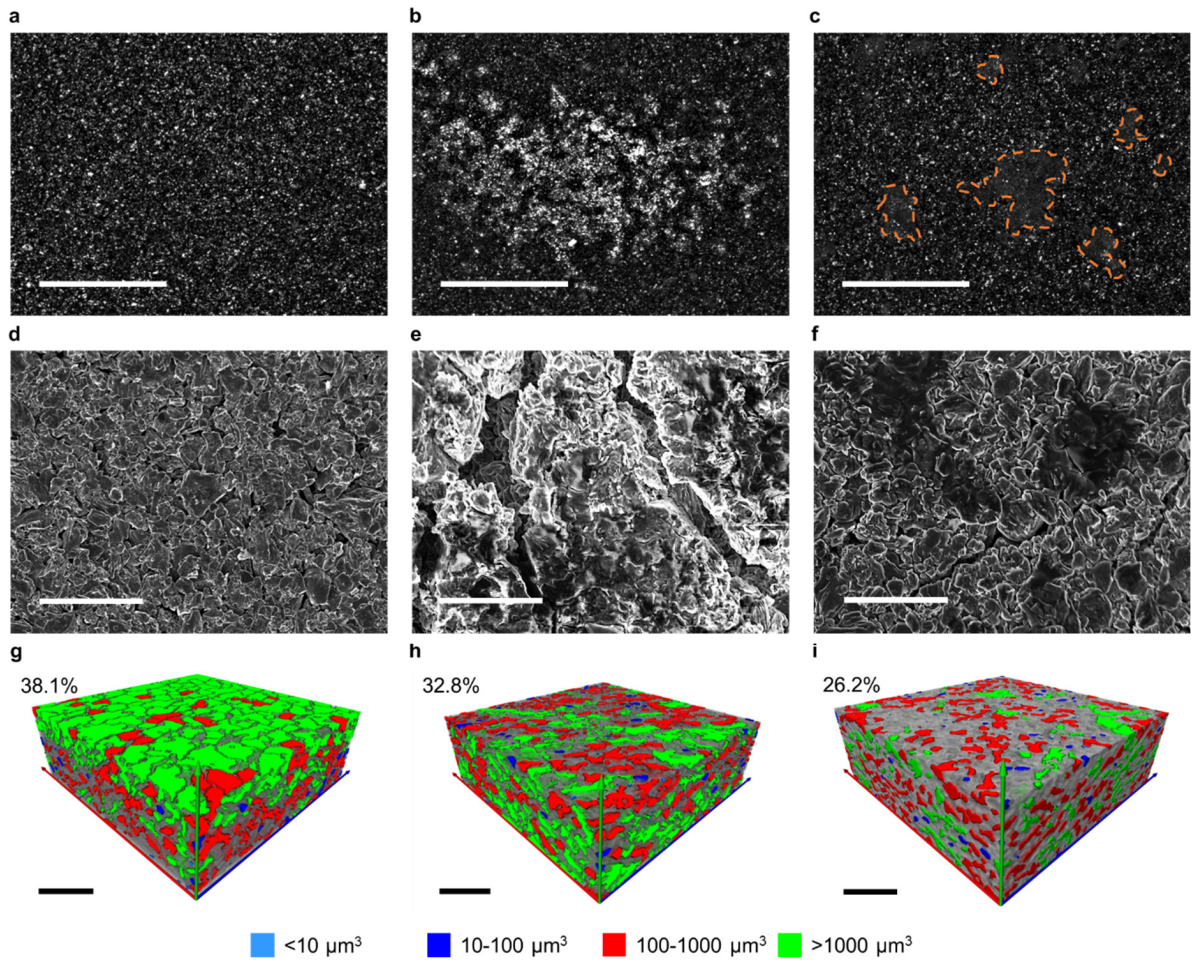

**Fig. 3 | Degradation mechanisms. a-c)** Optical images, **d-f)** SEM images, and **g-i)** tomography of pristine anode, and aged anode in the case of cooling and switch, respectively (see text for discussion). From the tomography, the reduction of pore size on the anode surface (*e.g.*, less large pores with volume > 1000 μm$^3$) indicates pore clogging. These postmortem characterizations of 5Ah LCO/C cells agree with the d$V$/d$t$ analysis and confirm that lithium plating induced degradation is mitigated due to the boosted battery kinetics at elevated temperature. The scale bar is 500 μm, 50 μm, and 50 μm in panels **a-c**, **d-f**, and **g-i**, respectively.

The proof-of-concept study clearly demonstrates the benefits of TMCP for XFC. To integrate our approach in existing BTMSs, the desirable design rules for ATS devices are: 1) switching ratio ≥ 10; 2) minimal impact on the system-level volumetric and gravimetric energy density; 3) small power consumption for switching; 4) zero power consumption in the ON/OFF state to maximize the energy efficiency; 5) compatibility with existing BTMSs. For the switching ratio ≥ 10, solid/solid mechanical switch based on contact and separation is a promising option based on a thorough review study of thermal switches by Wehmeyer *et. al.*[33]. Thus, we built a prototype device by integrating a mechanical thermal switch based on shape memory alloy (SMA) with a heat sink plate (Fig. 4a). SMA wires are selected for the active actuation using temperature-responsive phase and volume change, which satisfies the traits mentioned above. We purposely chose Nitinol SMA wires with a phase transition temperature > 60 °C to avoid any passive response to the environmental temperature. A bistable structure consisting of one spring steel strip and two pivot blocks is used for energy saving, *i.e.*, energy is only consumed for the change of state (Fig. 4a). The strip becomes concave or convex depending on the contraction of the SMA wire which triggers the rotation of pivot blocks. To switch from OFF to ON, a current pulse (1.5A and 5s) is applied to the right SMA wire to trigger the wire contraction and rotate the corresponding pivot block anticlockwise, and the gap between the battery and heat sink plate is closed as the spring strip becomes concave (Supplementary Fig. 11a). In the ON state, the pressure between the battery and heat sink comes from the elastic bands, which are used to mimic the pressure in practical battery packs. Reversely, heating the other SMA wire opens the gap (~0.5 mm) in a similar manner and brings the state back to OFF (Supplementary Fig. 11b). The electrical energy consumed for the change of thermal state is ~36 J per cycle, which is negligible (~0.05%) compared to the charge/discharge energy of the cells (~18.5 Wh).

Using this device, we perform the XFC cycling test of 5Ah LCO/C cells with the proposed TMCP. Infrared thermal images of the cell display the different battery surface temperature during XFC (42.2 °C) and discharging (30.4 °C) (Fig. 4a). ATS by the SMA device and the linear actuator demonstrates comparable thermal switching capability and results in similar evolution of temperature (Fig. 4b) and C rate (Fig. 4c) in a XFC cycle. Regardless of the ATS method, our TMCP leads to the XFC performance exceeding the targets set by DOE in terms of charge time (Fig. 4d), capacity retention (Fig. 4e), and discharge energy

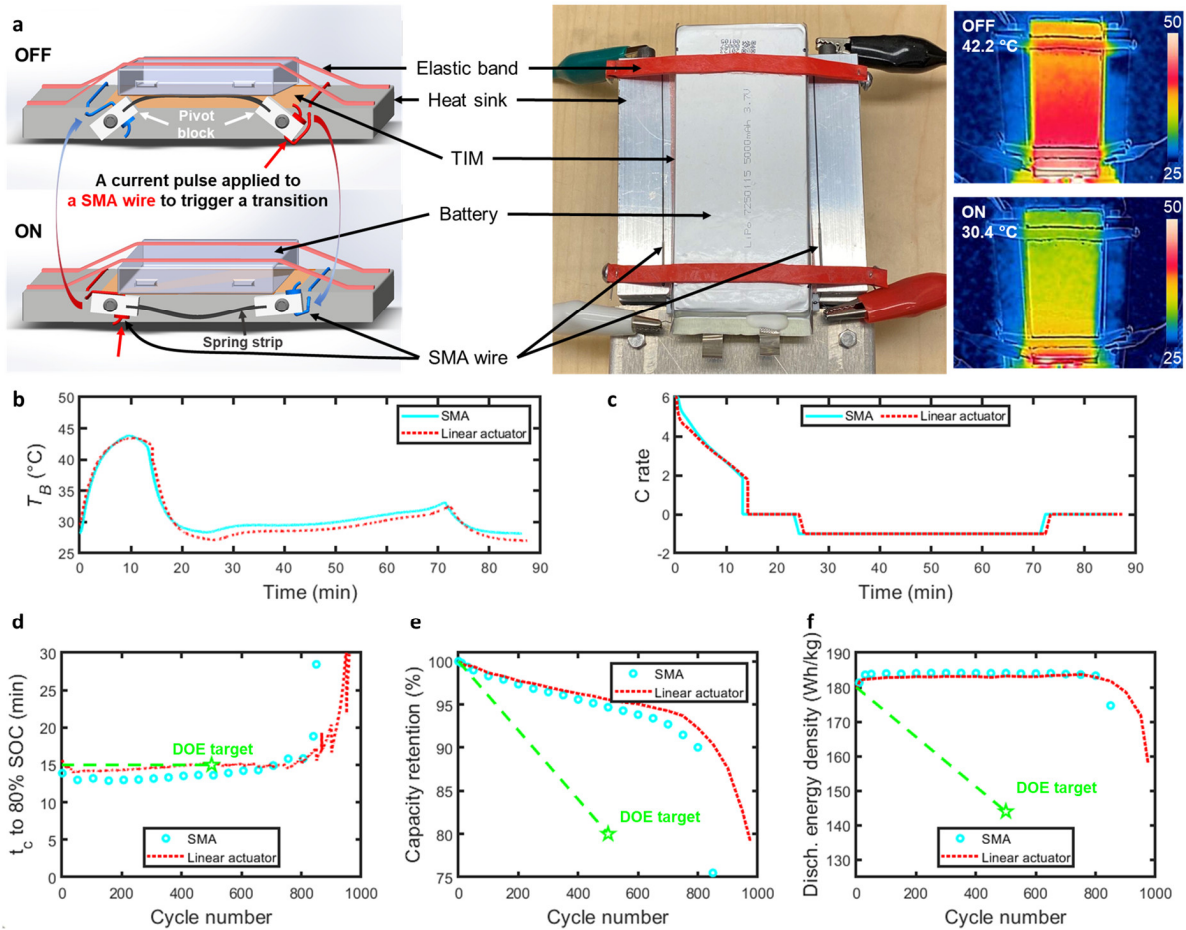

**Fig. 4 | Performance of SMA based thermal switch integrated with BTMSs. a,** Schematic, photo and infrared thermal image of the SMA thermal switch. The thermal contact is changed by heating the SMA wire with a current pulse, *i.e.*, the heated SMA wire contracts and triggers the switch. Representative evolution of **b**) temperature and **c**) C rate in a XFC cycle for 5Ah LCO/C cells by thermal switching. A comparison of **d**) charge time, **e**) capacity retention, and **f**) discharge energy density with the XFC target set by US DOE (the green dash line in **d-f**), *i.e.*, <15 min charge time, <20% capacity loss in 500 XFC cycles, and >180 Wh/kg discharge energy density after XFC, respectively.

density after XFC (Fig. 4f). The XFC cycle life (773) exceeds the DOE target (500) by 54.6% and the C/3 energy density after XFC is higher than the DOE target in the whole XFC cycle life. In addition to the XFC performance, the relative mass, volume, and material cost of the SMA-based thermal switch, compared to that of battery, is estimated to be 1.4%, 3.0%, 0.7-1.9%, respectively (Supplementary Note 3,

Supplementary Table 3 and 4). The mass, volume, and cost of this BTMS-integrated switch can be further reduced by optimizing the design in the future work. Considering the long R&D and commercialization cycles (~15-20 years) for new electrolytes and electrode materials, our approach using existing cost-effective materials (SMA) could provide a short to medium-term solution for enabling XFC. Besides XFC applications, ATS can improve battery performance at different temperatures by retaining or dissipating battery heat as needed, *e.g.*, retaining battery heat is beneficial for discharging at low temperature as demonstrated by Hao *et. al.*[34].

In summary, we have developed a thermal solution for enabling XFC in commercial high-energy density LIBs. Unlike previous innovations on battery materials, our approach leverages battery heat to improve the XFC performance using a BTMS-integrated thermal switch based on existing cost-effective materials. Considering the dependence of the optimal system temperature on the operating condition, the optimal temperature can be continuously adjusted depending on the condition of the cell by the ATS in a smart BTMS.

**Online content**

Any methods, additional references, Nature Research reporting summaries, source data, extended data, supplementary information, acknowledgements, peer review information; details of author contributions and competing interests; and statements of data and code availability are available.

**Methods**

**XFC cycling experiments**

Commercial high-energy 5Ah LCO/C LIBs (model number: PL-7250115-2C) were used in this work. 1C is the current needed to fully charge or discharge the nominal capacity (5Ah) in an hour. The nominal energy density and the calibrated C/3 energy density is 205.5 Wh/kg and 240.8 Wh/kg, respectively. From the manufacturer[35], the recommended maximum charge rate is 1C for such high-energy density pouch cells. The capacity of 5Ah was chosen for 6C XFC using our channel with a maximum current of 30 A. The charge cut-off voltage is 4.2 V and the discharge cut-off voltage is 3.0 V.

In the XFC cycling tests, batteries are charged at 6C to 80% SOC using a standard constant-current constant-voltage (CCCV) charging profile with a cutoff voltage of 4.2 V. After a 10 min rest, batteries are discharged at 1C to the cutoff voltage specified by the manufacturer. The rest time after discharge is 30 min for the case of insulation and 15 min for other cases, which ensures an approximate thermal equilibrium at the end of each XFC cycle, *e.g.*, <1 °C temperature rise compared to the ambient temperature. After each 50 XFC cycles, the capacity is calibrated by charging and discharging at C/3, and then the discharge energy density after XFC is quantified by 6C CCCV charging to 80% SOC and discharging at C/3.

All cells were tested with an 8-channel Arbin Laboratory Battery Testing System (LBT21084) in a TestEquity thermoelectric temperature chamber (TEC1). The cycler was calibrated by the manufacturer before use. The temperature chamber is off for the XFC cycling tests at 25±5 °C to mimic the real-world scenarios. It is set to 40 oC for simulating higher ambient temperature

**Thermal switch experiments**

For the conceptual thermal switch, a linear actuator (L12-R micro linear servos for RC & Arduino, Actuonix) was used and controlled with a microcontroller board (Arduino UNO). The board acquires the battery operation status (*i.e.*, charge or discharge) from the cell voltage reading and changes the status of actuator (*i.e.*, the original or elongated state) accordingly. The same logic is applied to the thermal switch device based on SMA. At the start or end of XFC, the status of switch is tuned by heating a specific SMA wire (0.254 mm Nitinol wire, Kellogg) using a current pulse. The thermal interface material (TIM) layer (Laird TPLI210) was selected for ensuring good thermal contact and durability. Infrared images were captured using a thermal imaging camera (Model number: FLIR E4).

**Post-mortem characterization**

The pristine and aged 5Ah LCO/C cells were fully discharged at C/3 to 2.75 V and then disassembled in a glove box. The graphite anodes were sealed in a chamber to ensure limited exposure to oxygen or moisture during optical characterizations using a confocal microscope (Lasertec L7 Hybrid). For SEM imaging and tomography, the electrode samples were washed multiple times with ethylmethyl carbonated solvent and dried in a vacuum chamber before the transfer to the instrument. SEM images were taken using FEI Helios 600 SEM/FIB. X-ray tomography was conducted using Beamline 8.3.2 at the Advanced Light Source (ALS) at Lawrence Berkeley National Laboratory. Details on the experiments, 3D reconstructions, and visualizations can be found in our prior work[36].

**Data availability**

The data supporting the findings of this study are available from the corresponding author on reasonable request.

**Acknowledgments** The authors acknowledge the support by Energy Efficiency and Renewable Energy, Vehicle Technologies Program, of the US Department of Energy under contract no. DEAC0205CH11231. This work used Beamline 8.3.2 at ALS, a DOE Office of Science User Facility under contract no. DEAC0205CH11231.

**Author contributions** Y. Z. and R. S. P. conceived the idea. Y. Z. and B. Z. designed and conducted the electrochemical simulations and cycling experiments. Y. Z., B. Z., Y. F., F. S., D. C., and Q. Y. conducted the postmortem characterizations. Y. Z., R. M., and R. S. P. designed the bistable thermal switch device. All the authors discussed the results. Y. Z., B. Z., and R. S. P. contributed to writing.

**Competing interests** Authors declare no competing interests.

**Additional information** Correspondence and requests for materials should be addressed to R.S.P.

**Supplementary Information**

**Extreme fast charging of batteries using thermal switching and self-heating**

Yuqiang Zeng[1,†], Buyi Zhang[1,2,†], Yanbao Fu[1], Fengyu Shen[1], Qiye Zheng[1,2], Divya Chalise[1,2], Ruijiao Miao[1,2], Sumanjeet Kaur[1], Sean D. Lubner[1], Michael C. Tucker[1], Vince Battaglia[1], Chris Dames[1,2], and Ravi S. Prasher[1,2,*]

[1]Energy Storage and Distributed Resources Division, Lawrence Berkeley National Laboratory, Berkeley, CA, 94720, USA

[2]Department of Mechanical Engineering, University of California, Berkeley, CA, 94720, USA

[†]: These authors contributed equally.

[*]: Author to whom correspondence should be addressed.

Email: rsprasher@lbl.gov

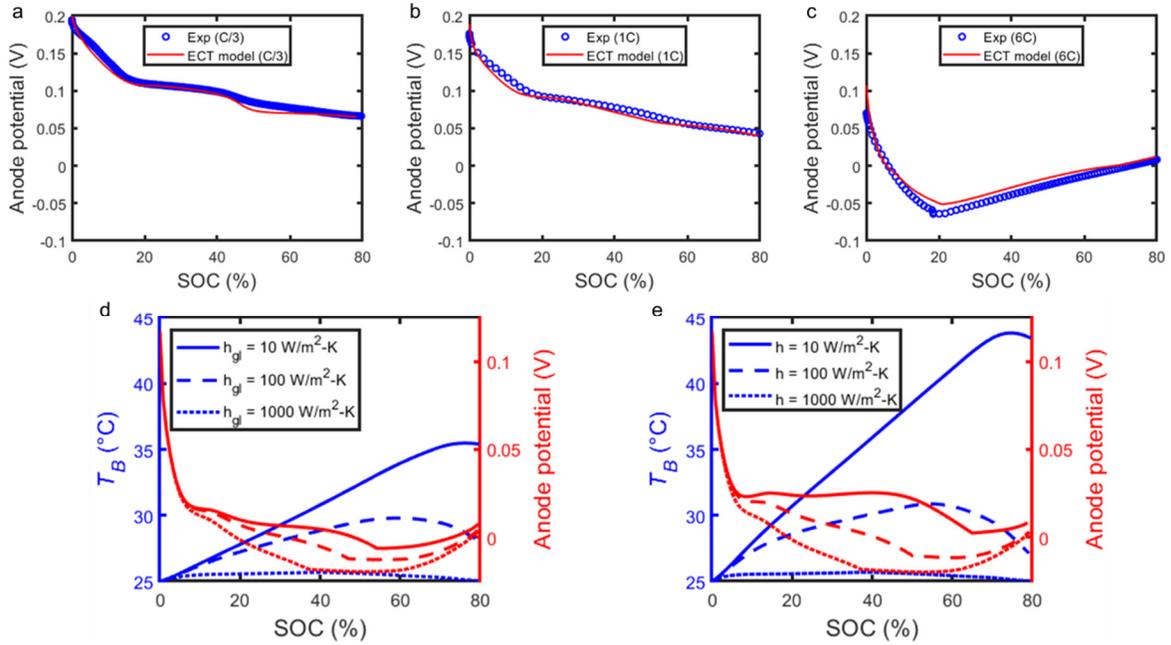

Supplementary Fig. 1 | Evaluation of thermal strategy using ECT model. Verification of our ECT model by comparing to the anode potential measured in a three-electrode cell during charging at a) C/3, b) 1C, and b) 6C. Evolution of battery temperature and anode potential during XFC of LCO/C cells using d) the system-level thermal strategy and e) the local thermal control of the cell. We use the global thermal conductance per unit area ($h_{gl}$) ranging from 10 W/m$^2$-K to 1000 W/m$^2$-K to mimic the different status of coolant flow. Regardless of the flow status, the heat leakage from the cell to the BTMS reduces the battery temperature rise during XFC and results in the negative anode potential. In contrast, the minimum anode potential can maintain positive by local thermal control of the cell with $h$ ~10 W/m$^2$-K.

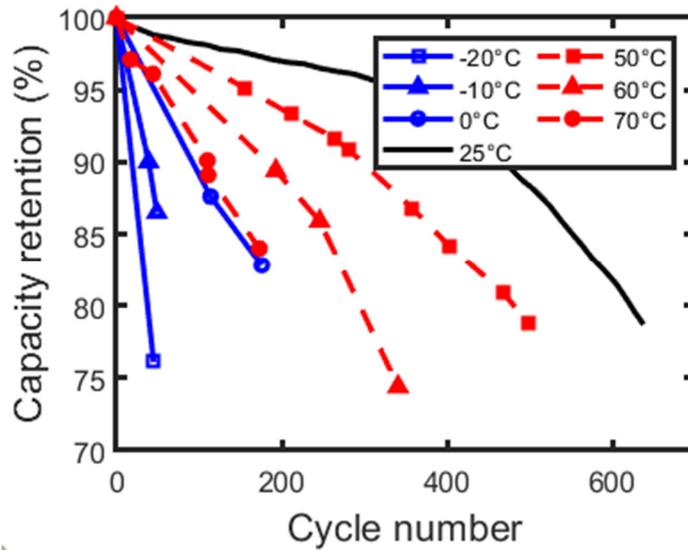

Supplementary Fig. 2 | Impact of operation temperature on the cycle life of LIBs[1]. The optimal operation temperature is ~25 °C for cycling at slow or medium rates (*e.g.*, 1C1C cycling). Increasing the temperature leads to increased side reactions and reduced cycle life, while the cycle life decreases at low temperature due to lithium plating. In the case of insulation average temperature is much higher than the case of thermal switch which adversely affects the cycle life of the insulation case.

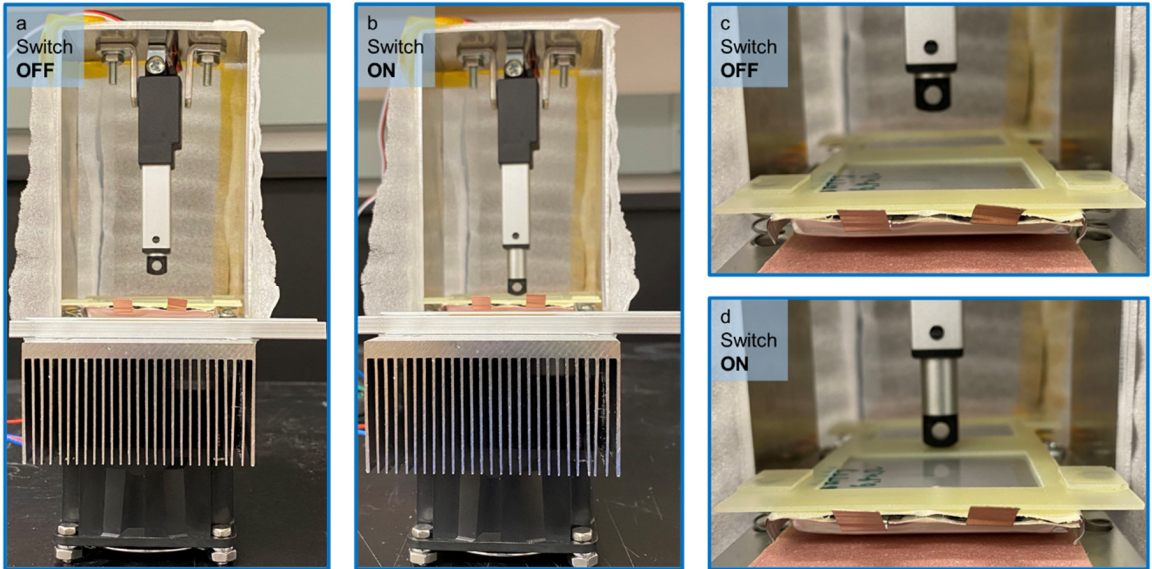

Supplementary Fig. 3 | Photo of the linear actuator used for ATS in the proof-of-concept study. The actuator in the original or elongated state corresponds to the switch a) OFF or b)ON, respectively. c) Switch OFF. A gap exists between the battery and heat sink and minimizes the heat transfer in between. d) Switch ON. The gap is closed as the actuator elongates and pushes the battery in contact with the heat sink.

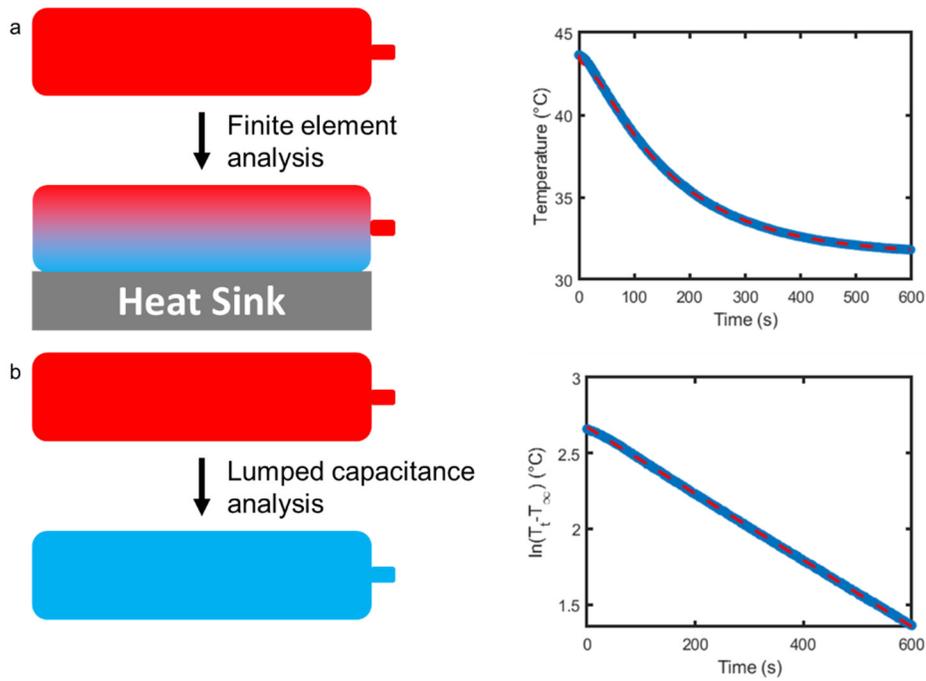

Supplementary Fig. 4 | Thermal analysis. a. Finite element analysis for determining the effective heat transfer coefficient ($h$) in the case of cooling and switch ON. The battery is heated in a thermal insulation state to a relatively uniform temperature and then set to a certain thermal condition (cooling or switch ON). The transient temperature during cooling is used for extracting the $h$. B) Lumped capacitance analysis for extracting the $h$ in the case of insulation and switch OFF by fitting to the transient temperature during resting.

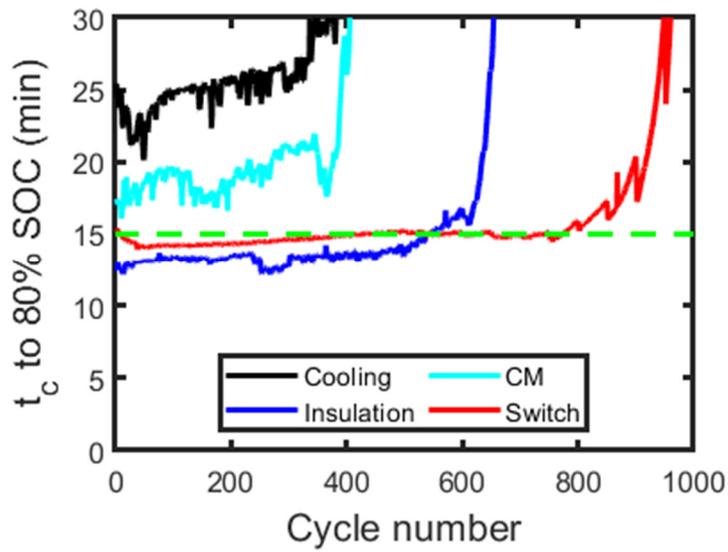

Supplementary Fig. 5 | Evolution of charge time with cycle number. The charge time to 80% SOC can increase to >15 min before losing 20% capacity due to battery aging.

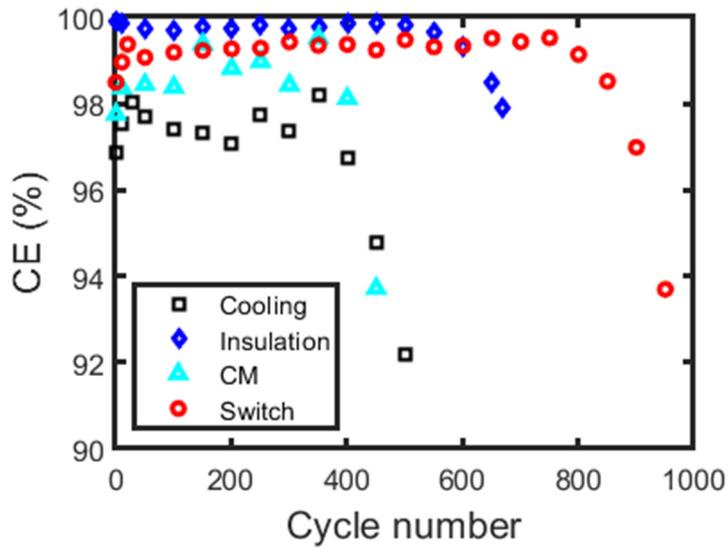

Supplementary Fig. 6 | Evolution of coulombic efficiency (CE) with cycle number. The low CE in the case of cooling and CM is attributed to the severe lithium plating related to XFC. The increased CE in the insulation and switch case owes to the mitigation of lithium plating at high temperature during XFC.

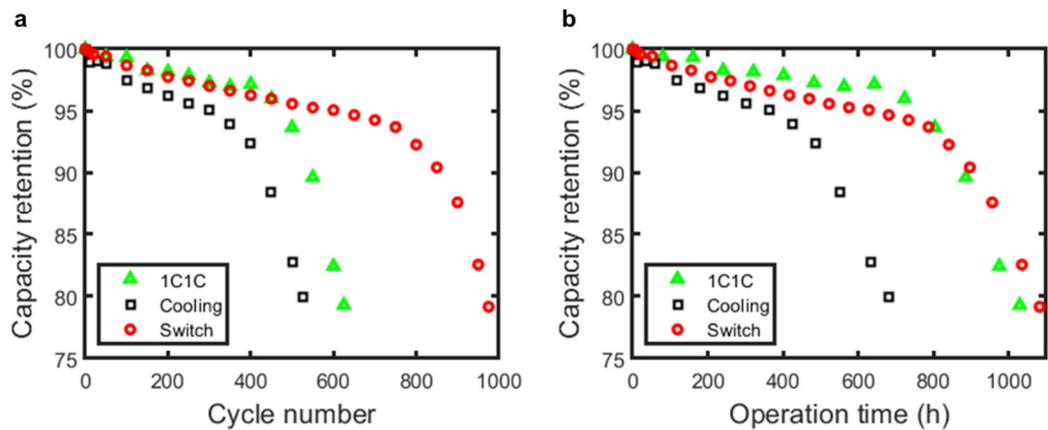

Supplementary Figure 7 | Comparison with 1C1C cycling results. Capacity retention as a function of a) cycle number and b) operation time for 5Ah LCO/C cells. The comparable operation time of 6C1C and 1C1C cycling suggests that the degradation related to XFC is largely reduced with our TMCP.

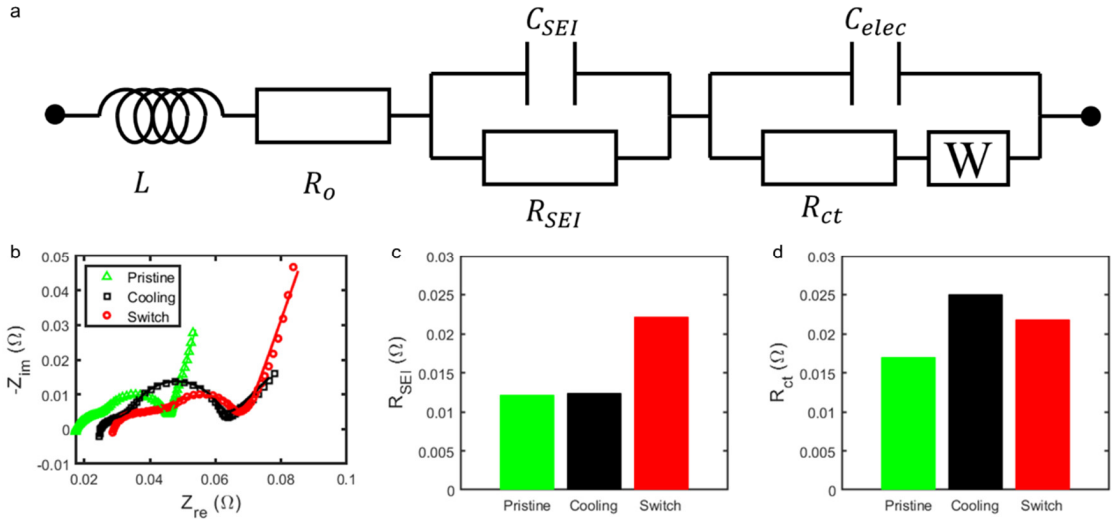

Supplementary Fig. 8 | EIS analysis of pristine and aged cells (5Ah LCO/C). a. The equivalent circuit model[2] used for fitting the electrochemical impedance spectra (EIS): $R_o$, $R_{SEI}$, and $R_{ct}$ is the ohmic resistance, solid-electrolyte interfacial (SEI) layer resistance, and the charge transfer resistance, respectively. $C_{SEI}$, $C_{elec}$, and $W$ represents the capacity of the SEI layer, the double layer capacitance, and the Warburg diffusion element, respectively. b. The EIS spectra of pristine and aged 5Ah LCO/C cells. From the analysis, the aged cell in the case of cooling has the higher $R_{ct}$ related to the severe lithium plating, and the highest $R_{SEI}$ is observed in the case of switch associated with the longest XFC cycle life. Comparison of c) $R_{SEI}$ and d) $R_{ct}$ in the aged cell from "cooling" and "switch" indicates the different aging mechanism: the larger $R_{SEI}$ in the switch case corresponds to more side reaction products and the more significant increase of $R_{ct}$ in "cooling" relates to the severe lithium plating.

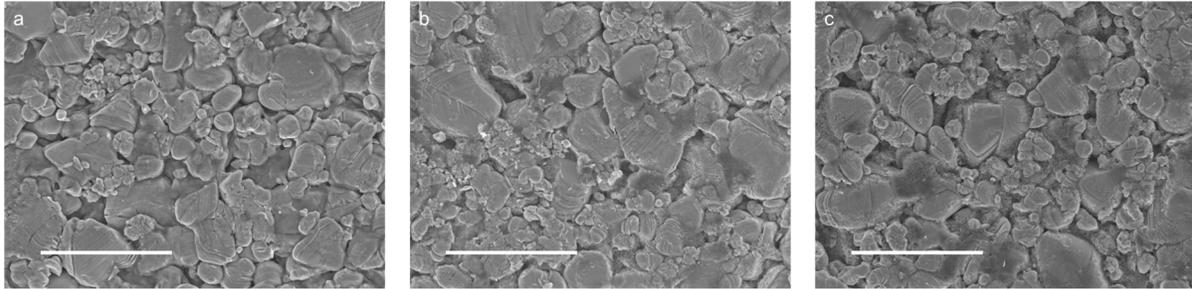

Supplementary Fig. 9 | SEM images of a) pristine cathode and aged cathodes for the case of cooling and switch. No observable cracking is found in the aged cathodes, which verifies the minimal effect of cathode aging on the capacity degradation in our study[3]. The scale bar is 25 μm in these panels.

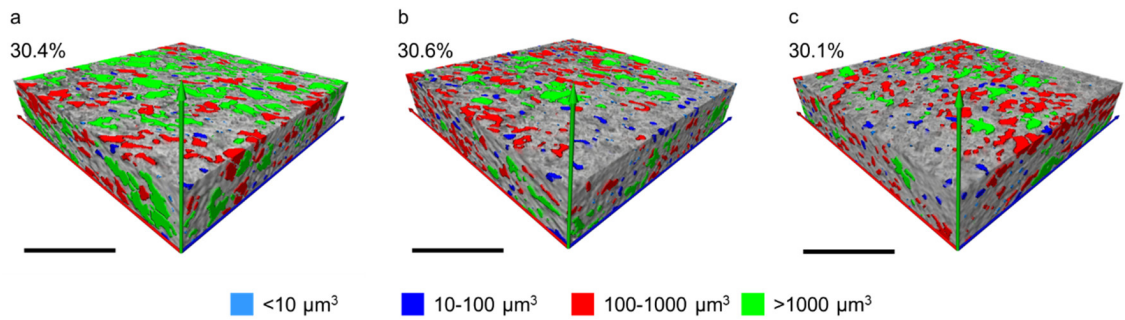

Supplementary Fig. 10 | Tomography of a) pristine cathode and aged cathodes for the case of cooling and switch (5Ah LCO/C cells). The cathode porosity remains almost the same as opposite to the observable change of anode porosity with battery aging. The scale bar is 100 μm in these panels.

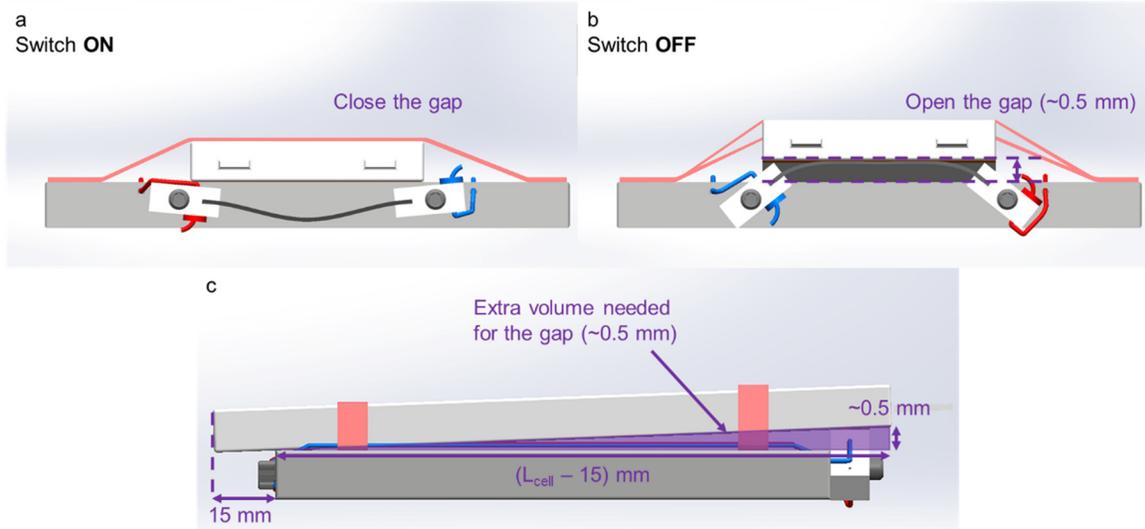

Supplementary Fig. 11 | Different views of the SMA-based thermal switch. Front view of a) Switch ON and b) Switch OFF shows the gap used for ATS. c. Side view of the switch shows the volume added to BTMSs due to the presence of gap.

Supplementary Table 1 | Parameters from the EIS analysis of pristine and aged 5Ah LCO/C cells

|  | $L$ | $R_0$ | $R_{SEI}$ | $R_{ct}$ | $C_{SEI}$ | $C_{elec}$ | $W$ | $n_{SEI}$ | $n_{elec}$ | $n_W$ |
|---|---|---|---|---|---|---|---|---|---|---|
| Pristine | -7.08273 | 0.016787 | 0.012135 | 0.016909 | 0.540297 | 0.826039 | -2.75266 | 0.605933 | 1 | 0.817654 |
| Cooling | -7.14281 | 0.023981 | 0.012327 | 0.025 | 0.355592 | 0.485631 | -2.40879 | 0.562289 | 0.997486 | 0.488628 |
| Switch | -7.06626 | 0.025389 | 0.022153 | 0.021727 | 0.043622 | 0.644885 | -2.47911 | 0.498406 | 0.834558 | 0.77722 |

Note: SI units are used for all variables.

Supplementary Table 2 | Capacity loading of pristine electrodes and aged electrodes in the switch case for 5Ah LCO/C cells

| Calibration at C/10 | Pristine Anode | Pristine Cathode | Cycled Anode | | Cycled Cathode | |
|---|---|---|---|---|---|---|
|  |  |  | Edge | Center | Edge | Center |
| $Q_c$ (mAh) | 3.92 | 3.62 | 3.82 | 3.97 | 3.56 | 3.68 |
| $Q_d$ (mAh) | 3.91 | 3.59 | 3.81 | 3.96 | 3.52 | 3.65 |
| Loading (mAh/cm$^2$) | 3.09 | 2.83 | 3.01 | 3.12 | 2.78 | 2.88 |

Supplementary Table 3 | Mass, material cost, and volume of the SMA switch and LIBs

|  | Mass (g) | Unit cost | Material cost ($) | L×H$^*$ (mm) |
|---|---|---|---|---|
| SMA wires | 0.08 | $30-300 /kg[4] | 0.0024-0.024 | - |
| Pivot blocks | 0.64 | $15 /kg[5] | 0.0096 | - |
| Spring steel strip | 0.44 | $0.25-2.70 /kg[6] | 0.00011-0.0012 | - |
| Total | 1.16 | - | 0.01211-0.0348 | ($L_{cell}$-15)×~0.5 |
| 5Ah LCO/C | 82.36 | $97 /kWh[7] | 1.7945 | 117×7.2 |

Supplementary Table 4 | Relative mass, material cost, and volume of the SMA switch

|  | Relative mass (%) | Relative material cost (%) | Relative volume (%) |
|---|---|---|---|
| 5Ah LCO/C | 1.4 | 0.7-1.9 | ~3.0 |

Supplementary Note 1 | Electrochemical-thermal (ECT) simulation

Electrochemical-thermal (ECT) simulations were performed in COMSOL Multiphysics 5.6. We coupled the Lithium-Ion Battery Module and the Heat Transfer Module for the simulation of battery cycling in different thermal conditions, based on a Newman pseudo 2D electrochemical model and a lumped thermal model. To verify our ECT model, we simulated the anode potential during XFC and observed a good agreement with experiment results from a three-electrode cell (see Supplementary Figs. 1a-c). In this verification study, we assembled three-electrode cells using materials as described in our prior work[8].

For the ECT simulation of a representative commercial battery, *e.g.*, commercial LCO/C cells , the thermal model uses the heat generation rate form the electrochemical simulation as input, and the evolution of battery temperature during charging in different thermal conditions is calculated. The battery with or without BTMS is treated as a lumped thermal system. This simplified treatment is appropriate low heat transfer coefficients are more important for our study. In the simulation of cell packs including BTMS, the BTMS was considered as an additional thermal mass in the lumped thermal system. According to Yang *et. al.*[9], the gravimetric cell-to-pack ratio is 55-65%, which means 35-45% of the pack weight is taken by management system, metals, cabling and others. We estimate the heat capacity of management system as the average of copper and coolants (water/glycol)[10].

Gen2 electrolyte (1.2 M LiPF$_6$ in EC:EMC 3:7) was used in all the studies, which is the baseline electrolyte for XFC suggested by the US DOE. We use the transport properties of electrolyte, graphite and LiCoO$_2$, *e.g.*, diffusion coefficient and conductivity, provided by COMSOL material library. The open circuit voltage (OCV) of graphite and LiCoO$_2$ are also from COMSOL material library.

Electrode

| Parameter | Anode (Graphite) | Cathode (LiCoO$_2$) |
|---|---|---|
| Thickness ($\mu m$) | 55.4 | 52.2 |
| Initial Porosity | 0.382 | 0.356 |
| Loading ($mAh/cm^2$) | 2.34 | 2.11 |
| Particle radius ($\mu m$) | 5 | 5[11] |
| Specific surface area ($m^{-1}$) | $3.267 \times 10^5$ | $2.97 \times 10^5$ |
| Usable capacity range by lithium intercalation fraction | 0.12~0.95 | 0.453~0.994 |
| Bruggeman exponent, p | 2.55 | 2.2 |
| Exchange current density ($A/m^2$) | 2.1[11] | 2.1 |
| Activation energy of exchange current density ($kJ/mol$) | 68[11] | 69[12] |
| Solid-state diffusivity, $D_s (cm^2/s)$ | COMSOL Materials Library | COMSOL Materials Library |
| Activation energy of solid-state diffusivity ($kJ/mol$) | | 25[13] |

Separator

| Electrolyte concentration (mol/L) | 1.2 |
|---|---|
| Thickness ($\mu m$) | 25 |
| Porosity | 0.41 |
| Bruggeman exponent, p | 2 |

Cell

| Cathode Material | $LiCoO_2$ |
|---|---|
| Specific heat of cell, $c_p$ ($J/(kg*K)$) | 1100[14] |
| Mass of the battery ($kg$) | 0.211 |
| Thickness ($mm$) | 10 |

Supplementary Note 2 | Thermal Analysis

Effective thermal conductance per unit area ($h$) was obtained from the experiments by matching the temperature profile (Fig. 1e) from the experiment and detailed thermal model in COMSOL.

Supplementary Note 3 | Estimation of mass, volume, and material cost for the SMA switch

Each SMA device, as shown in Fig. 4a, consists of 2 SMA wires (Nitinol alloy), 1 spring steel strip (1095 spring steel), and 2 pivot blocks (Teflon). Supplementary Table 3 summarizes the mass, volume, and material cost of each component. For the estimation of volume, the volume of the strip and pivot blocks is not regarded as extra as they are built on the sidewall of the heat sink plate (*i.e.*, unusable space in existing BTMSs). For this reason, only the volume related to the gap that affects the cell-to-pack volumetric ratio is considered in the analysis (Supplementary Fig. 11c). We estimate the relative mass, volume, and material cost the SMA switch compared to that of batteries (Supplementary Table 4). Compared to 5Ah LCO/C cells, the relative mass, volume, and material cost of the SMA device is 1.4%, ~3.0%, and 0.7-1.9%, respectively. These ratios could reduce as a battery of higher capacity is used.